\documentclass[a4paper,pra,reprint, twocolumn,superscriptaddress]{revtex4}

\usepackage{ulem}
\usepackage{amssymb}
\usepackage{amsmath}
\usepackage{epsfig}
\usepackage{color}
\usepackage{graphics, graphicx}
\usepackage{bbold}
\usepackage{psfrag}
\usepackage{mathcomp}
\usepackage{amsmath}
\usepackage{amssymb}
\usepackage{mathrsfs}
\usepackage{subfigure}
\usepackage{verbatim}
\usepackage{xcolor}
\usepackage[colorlinks,citecolor=blue,urlcolor=blue]{hyperref}

\usepackage{mathrsfs}

\begin{document}

	\date{\today}
	\title{Spontaneous formation of polar superfluid droplets in a p-wave interacting Bose gas}
	
	\author{Zehan Li}
	\affiliation{Department of Physics and Astronomy, University of Pittsburgh, Pittsburgh, PA 15260, USA}
	
	\author{Jian-Song Pan}
	\email{panjsong@sjtu.edu.cn}
	\affiliation{Wilczek Quantum Center, School of Physics and Astronomy and T. D. Lee Institute,
		Shanghai Jiao Tong University, Shanghai 200240, China}
	
	\author{W. Vincent Liu}
	\email{wvliu@pitt.edu}
	\affiliation{Department of Physics and Astronomy, University of Pittsburgh, Pittsburgh, PA 15260, USA}
	\affiliation{Wilczek Quantum Center, School of Physics and Astronomy and T. D. Lee Institute, Shanghai Jiao Tong University, Shanghai 200240, China}
	\affiliation{Shenzhen Institute for Quantum Science and Engineering and Department of Physics, Southern University of Science and Technology, Shenzhen 518055, China}
	
	\begin{abstract}
		We study the quantum fluctuations in the condensates of a mixture of bosonic atoms and molecules with interspecies p-wave interaction. Our analysis shows that the quantum phase of coexisting atomic and molecular condensates is unstable at the mean-field level. Unlike the mixture of s-wave interaction, the Lee-Huang-Yang correction of p-wave interaction is unexpectedly found here to exhibit an opposite sign with respect to its mean-field term above a critical particle density. This quantum correction to the mean-field energy provides a remarkable mechanism to self-stabilize the phase. The order parameter of this superfluid phase carries opposite finite momenta for the two atomic species while the molecular component is a polar condensate. Such a correlated order spontaneously breaks a rich set of global U(1) gauge, atomic spin, spatial rotation and translation, and time-reversal symmetries. For potential experimental observation, the phenomenon of anisotropic polar superfluid droplets is predicted to occur, when the particle number is kept finite.
	\end{abstract}
	\pacs{67.85.Lm, 03.75.Ss, 05.30.Fk}

	\maketitle
	
	\section{Introduction}
	Quantum fluctuation is one of the most intrinsic properties of quantum mechanics, which is responsible for many fascinating physical phenomena, such as Casimir effect and abundant quantum phase transitions. Recently, Petrov showed that quantum fluctuation reflected by Lee-Huang-Yang (LHY) correction can prevent a mean-field-unstable Bose gas from collapsing~\cite{petrov2015quantum}. The competition between the mean-field attraction and LHY repulsion stabilizes the Bose gas into a self-bound liquidlike droplet state. Subsequently, several experimental groups reported this novel quantum state with the prediction of Petrov~\cite{semeghini2018self,cheiney2018bright,cabrera2018quantum}. In order to protrude the action of LHY correction, which is typically small in the dilute limit, Petrov suggested to subtly balance the inter- and intra-species interactions at the mean-field level. Owing to its unique formation mechanism, the self-bound state shows many interesting features, such as the quantum droplet is self-trapped and evaporated without external potential~\cite{petrov2015quantum}.
	
	The properties of quantum droplet are linked to the properties of interaction between particles. It is natural to ask if quantum droplet can be stabilized with other types of interaction and what their properties might be. It was also found that quantum droplets can be stabilized in a dipolar Bose gas benefiting from the competition between the dipolar interaction and s-wave contact interaction~\cite{ferrier2016observation,kadau2016observing,chomaz2016quantum,schmitt2016self}. The quantum droplets in a dipolar Bose gas are anisotropic and form a regular array, as a consequence of the dipolar interaction is anisotropic and long-ranged. Morover, it is also predicted quantum droplets can be stabilized with the assistance of three-body interaction~\cite{bulgac2002dilute,sekino2018quantum} and spin-orbit coupling~\cite{cui2018spin}.
	
	Here we study the beyond-mean-field ground state of a p-wave interacting Bose gas, and predict the existence of finite-momentum anisotropic self-stabilized quantum droplet.  At the mean-field level, this p-wave interacting Bose gas typically has three ground-state phases: atomic superfluid (ASF) phase with only the atomic condensate, atomic-molecular superfluid (AMSF) phase with both atomic and molecular condensates, and molecular superfluid (MSF) phase with only the molecular condensate. We find AMSF phase is unstable and tends to collapse. Unlike pure s-wave interaction~\cite{LHYoringinalpaper}, we find the sign of the LHY correction of p-wave interaction may be different from that of the mean-field term when varying particle densities. A balance between the mean-field part and LHY correction exists for certain particle density, which gives rise to a self-stabilized (-bound) state without external potential. It is shown the self-stabilized state even survives in the dilute limit estimated with scattering volume. In addition to the U(1) global phase symmetry, the rotation, translation and time-reversal symmetries are found to be spontaneously broken by the presence of finite momentum of the order parameters. The result ground state is predicted to be an anisotropic quantum droplet with finite momentum for a system with finite particle number.
	
	\begin{table*}
		\begin{tabular}{|c|c|c|c|c|}
			\hline
			symmetry & $\hat{\psi}_1(\textbf{r})$ & $\hat{\psi}_2(\textbf{r})$ & $\hat{\phi}_{x,y,z}(\textbf{r})$ & $\boldsymbol{\nabla}$ \tabularnewline
			\hline
			\hline
			$U_N(1)$  & $e^{i\theta}\hat{\psi}_1$ & $e^{i\theta}\hat{\psi}_2$ & $e^{2i\theta}\hat{\phi}_{x,y,z}$ & $-$\tabularnewline
			\hline
			$[SU(2)/U_y(1)]$  & \multicolumn{2}{c}{$e^{i\theta_x \sigma_x+i\theta_z \sigma_z}(\hat{\psi}_1, \hat{\psi}_2)^T$} & $-$ & $-$ \tabularnewline
			\hline
			$SO(3)$ & $-$ & $-$ & $e^{i\sum_{i=x,y,z}\overline{\theta}_{i}\lambda_i}\hat{\phi}$ & $e^{i\sum_{i=x,y,z}\overline{\theta}_{i}\lambda_i}\boldsymbol{\nabla}$\tabularnewline
			\hline
			$Tr$ & $\hat{\psi}_1(\textbf{r}+\textbf{r}')$ & $\hat{\psi}_2(\textbf{r}+\textbf{r}')$ & $\hat{\phi}_{x,y,z}(\textbf{r}+\textbf{r}')$ & $-$ \tabularnewline
			\hline
			$\mathcal{T}$ & $\sum_{p_1}e^{-i\textbf{p}_1\cdot \textbf{r}} \hat{a}_{1,-\textbf{p}_1}$ & $\sum_{p_2}e^{-i\textbf{p}_2\cdot \textbf{r}} \hat{a}_{2,-\textbf{p}_2}$ & $e^{-i(\textbf{p}_1+\textbf{p}_2)\cdot \textbf{r}} \hat{b}_{i,-\textbf{p}_1-\textbf{p}_2}$ & $-$ \tabularnewline
			\hline
		\end{tabular}
		\caption{Symmetry transformation. $U_N(1)$: $\theta\in [0,2\pi)$ is an arbitrary angle. This symmetry correponds to the total number conservation. $[SU(2)/U_y(1)]$ with spin rotation symmetry $U_y(1)$ generated by $\sigma_y$: $\theta_x$ and $\theta_z$ are arbitrary angles. Here $\sigma_{x,y,z}$ are the Pauli matrices. $SO(3)$: $\lambda_{x,y,z}$ are defined in Eq.~(\ref{lambdamatrix}) and $\overline{\theta}_{x,y,z}$ are arbitrary rotation angles. $Tr$: $\textbf{r}'$ is an arbitrary displacement vector in 3D spatial coordinate. $\mathcal{T}$ time-reversal: We use momentum representation to expand $\hat{\psi}_1$ and $\hat{\psi}_2$ fields. Due to momentum conservation, the momentum of molecule fields is restricted to $\textbf{p}_1+\textbf{p}_2$.}\label{symmetry_table}
	\end{table*}
	
	\section{Model}
	Inspired by the experimental observations of p-wave Feshbach resonance in the mixture of $^{85}$Rb and $^{87}$Rb atoms~\cite{papp2008tunable,dong2016observation}, we consider a mixture of two distinguishable species of bosonic atoms respectively created by $\hat{\psi}_1^{\dagger}(\boldsymbol{r})$ and $\hat{\psi}_2^{\dagger}(\boldsymbol{r})$ with interspecies p-wave interaction. The p-wave interaction arises from a p-wave Feshbach resonance by coupling with three closed molecular channels denoted by $l_z=-1,0,1$. Here $l_z\hbar$ are the magnetic angular momentum carried by the molecules on the closed channels, which are created by $\hat{\phi}_{l_z=-1,0,1}^{\dagger}(\boldsymbol{r})$ respectively. It will be convenient to discuss the physics with bases $\hat{\phi}_{i=x,y,z}^{\dagger}$, which are related with $\hat{\phi}_{l_z=-1,0,1}^{\dagger}$ through $\phi_{\pm 1}^\dagger=(\phi_x^\dagger \pm i\phi_y^\dagger)/\sqrt{2}$, and $\phi_0^\dagger=\phi_z^\dagger$. To focus on the physics arising from p-wave interaction, we will restrict our attention on the case where the closed channels are degenerate and background (non-resonant) interactions are neglectable. The system we consider is characterized by Hamiltonian density
	\begin{equation}\label{eq_mainH}
	\begin{split}
	\mathcal{H}=&
	\sum_{\sigma=1,2}\left.\hat{\psi}_\sigma^\dagger(-\frac{\nabla^2}{2m})\hat{\psi}_\sigma\right.
	\left.+\sum_{i=x,y,z}\hat{\phi}_i^\dagger(-\frac{\nabla^2}{4m}-\epsilon_0)\hat{\phi}_i\right.\\
	&\left.+\sum_{i=x,y,z}[\frac{\overline{g}}{2}\hat{\phi}_i^\dagger (\hat{\psi}_1, \hat{\psi}_2) \sigma_y \partial_i (\hat{\psi}_1, \hat{\psi}_2)^T+h.c.\right.],
	\end{split}
	\end{equation}
	where the atomic masses have been assumed to be the same, i.e. $m_1=m_2=m$, $\epsilon_0$ is the detuning of molecule channels, $\bar{g}$ represents the strength of p-wave interaction, and $\sigma_y$ is the Pauli matrix.
	Here the reduced Plank constant $\hbar$ has been set as $1$.
	
	Our model possesses $U_N(1) \times [SU(2)/U_y(1)] \times SO(3) \times Tr \times \mathcal{T}$ symmetries, where $U_N(1)$ is the global gauge symmetry, $[SU(2)/U_y(1)]$ the spin rotation symmetry around $x$ and $z$ directions, $SO(3)$ the 3-dimensional spatial rotation symmetry, $Tr$ the translation symmetry in the absence of an external field, and $\mathcal{T}$ the time reversal symmetry. The symmetry transformations are listed in Tab.~\ref{symmetry_table}. It is worth noting that spin-rotation symmetry $[SU(2)/U_y(1)]$ is reduced to a spin-rotation symmetry $U_z(1)$ generated by $\sigma_z$ in presence of intraspecies s-wave interaction~\cite{radzihovsky2009p,pwave2011Sungsoo}.  In $SO(3)$ rotation symmetry, the atom fields are scalar fields, so they remain constant under $SO(3)$ transformation. However, molecular field $\hat{\phi}$ and gradient operator $\boldsymbol{\nabla}$ are all vector fields, and they are transformed by a 3D spatial rotation. In Tab.~\ref{symmetry_table}, the generators of rotation symmetry $\lambda_{x,y,z}$ are given by,
	\begin{equation}\label{lambdamatrix}
	\begin{split}
	&\lambda_x=\begin{pmatrix} 0 & 0 & 0 \\ 0 & 0 & -i \\ 0 & i & 0 \end{pmatrix},
	\lambda_y=\begin{pmatrix} 0 & 0 & i \\ 0 & 0 & 0 \\ -i & 0 & 0 \end{pmatrix},
	\lambda_z=\begin{pmatrix} 0 & -i & 0 \\ i & 0 & 0 \\ 0 & 0 & 0 \end{pmatrix}.
	\end{split}
	\end{equation}
	Time-reversal symmetry $\mathcal{T}$ is given by reversing the momentum of atomic and molecular field operators, i.e. transforming $\hat{a}_{1,\textbf{p}_1}, \hat{a}_{2,\textbf{p}_2}$, and $\hat{b}_{i,\textbf{p}_1+\textbf{p}_2}$ as $\hat{a}_{1,-\textbf{p}_1}, \hat{a}_{2,-\textbf{p}_2}$, and $\hat{b}_{i,-\textbf{p}_1-\textbf{p}_2}$, respectively.
	
	The total particle number $N$ and atomic number difference $\delta N$ are defined as below,
	\begin{equation}
	N_1+N_2+2N_M=N,\quad N_1-N_2=\delta N,
	\end{equation}
	where we use $N_{1,2}=\int d^3r\langle\hat{\psi}_{1,2}^{\dagger}\hat{\psi}_{1,2}\rangle$ and $N_M=\sum_{i=x,y,z}\int d^3r\langle\hat{\phi}^{\dagger}_i\hat{\phi}_i\rangle$ to denote the numbers of atoms and molecules, respectively. Here $\langle\cdots
	\rangle$ represents the average over the ground state. Obviously $N$ and $\delta N$ are conserved in our model, which correspond to the $U_N(1)$ and $[SU(2)/U_y(1)]$ symmetries.

	\section{Mean-Field Ground State}
	As the foundation of beyond-mean-field study, we need to characterize the ground state at the mean-field level at first. We use the mean fields $\Psi_1=\langle\hat{\psi}_1\rangle$, $\Psi_2=\langle\hat{\psi}_2\rangle$ and $\Phi_i=\langle\hat{\phi}_i\rangle$ to describe the atomic and molecular condensates. The mean-field ground state of a p-wave resonant Bose gas including considerable large intraspecies s-wave interaction has been systematically discussed before~\cite{radzihovsky2009p,pwave2011Sungsoo}. Three mean-field stable phases for the ground states: atomic (ASF), atomic-molecular (AMSF) and molecular (MSF) superfluid, are found. Typically, the atomic condensates carry finite momentum due to the p-wave interaction in AMSF phase. Actually, the ground-sate phase diagram of our model is similar to the case there. While due to the lack of intraspecies s-wave interaction (or due to weak intraspecies s-wave interaction), it is shown the ground states of our model may be unstable in the mean field level.
	
	\begin{table*}
		\small\addtolength{\tabcolsep}{-2pt}
		\begin{tabular}{|c|c|c|c|c|c|c|c|}
			\hline
			Phase & $\epsilon_0$ & $n_M$ & $n_1=n_2$ & $Q$ & $\mu$ & $\ Z \ $ & $E_0/V$  \tabularnewline
			\hline
			\hline
			ASF & $\epsilon_0<-\frac{1}{2}\overline{g}^2mn$ & $0$ & $\frac{1}{2}n$ & $0$ & $0$ & $0$ & $0$\tabularnewline
			\hline
			AMSF & $-\frac{1}{2}\overline{g}^2mn<\epsilon_0<\frac{1}{2}\overline{g}^2mn$ & $\frac{1}{4}n+\frac{\epsilon_0}{2\overline{g}^2m}$ & $\frac{1}{4}n-\frac{\epsilon_0}{2\overline{g}^2m}$ & $-\overline{g}m\sqrt{\frac{1}{4}n+\frac{\epsilon_0}{2\overline{g}^2m}}$ & $-\frac{1}{8}\overline{g}^2mn-\frac{1}{4}\epsilon_0$  & $0$ & $-\frac{1}{16\overline{g}^2m }(\overline{g}^2mn+2\epsilon_0)^2$\tabularnewline
			\hline
			MSF &  $\epsilon_0>\frac{1}{2}\overline{g}^2mn$ & $\frac{1}{2}n$ & $0$ & $-\frac{1}{\sqrt{2}}\overline{g}m\sqrt{n}$ & $-\frac{1}{4}\overline{g}^2mn$ & $0$ & $-\frac{1}{2}\epsilon_0 n$\tabularnewline
			\hline
		\end{tabular}
		\caption{Table of ground state phases. Here we have three phases by setting different detuning. ASF, AMSF and MSF are the atomic, atomic-molecular and molecular condensate phases, respectively. }\label{tab_PD}
	\end{table*}

	As the typical feature of p-wave interaction, the atomic condensates generally carry finite momentum due to the shift of energy minimum in momentum space by the interaction terms~\cite{radzihovsky2009p, pwave2011Sungsoo}. Although a general description of atomic order parameters should be written as
	$\Psi_{1}=\sum_{\textbf{Q}_n}\Psi_{1,\textbf{Q}_n}e^{-i\textbf{Q}_n \cdot \textbf{r}}$ and $\Psi_{2}=\sum_{\textbf{Q}_n}\Psi_{2,-\textbf{Q}_n}e^{i\textbf{Q}_n \cdot \textbf{r}}$, it is shown that the assumption $\textbf{Q}_n=\textbf{Q}$ is sufficient to capture the ground state in presence of intraspecies s-wave interaction~\cite{radzihovsky2009p, pwave2011Sungsoo}, i.e.
	\begin{equation}
	\Psi_{1}=\Psi_{1,Q}e^{-i\textbf{Q} \cdot \textbf{r}},\quad \Psi_{2}=\Psi_{2,-Q}e^{i\textbf{Q} \cdot \textbf{r}},
	\end{equation}
	due to the lack of $\sigma_x$ spin-rotation symmetry for atomic components. Correspondingly the molecular components are space-independent, since the molecular fields only feel a homogeneous potential by atoms. Considering the symmetries of our model, we have the following ground-state ansatz
	\begin{equation}\label{eq_ouransatz}
	\begin{split}
	&\Psi=\sqrt{n_{A}}e^{i\theta}e^{i(\theta_x\sigma_x+\theta_{z}\sigma_z)}\begin{pmatrix} \cos{\chi_A}e^{-i\textbf{Q} \cdot \textbf{r}} \\ \sin{\chi_A}e^{i\textbf{Q} \cdot \textbf{r}} \end{pmatrix},\\
	&\textbf{Q}=e^{i\sum_{i=x,y,z}\overline{\theta}_{i}\lambda_i}\textbf{Q}_0,\\
	&\Phi=\sqrt{n_M}e^{i(2\theta+\theta_M)}e^{i\sum_{i=x,y,z}\overline{\theta}_{i}\lambda_i} \begin{pmatrix} \cos{\chi_M} \\ i \sin{\chi_M} \\ 0 \end{pmatrix},
	\end{split}
	\end{equation}
	\noindent
	where $\theta, \theta_M \in [0,2\pi)$ are $U(1)$ phases, $\theta_{x,z} \in [0,2\pi)$ are $[SU(2)/U_y(1)]$ spin rotation angles, $\overline{\theta}_{x,y,z}$ are $SO(3)$ rotation angles, $\chi_A,\chi_M\in[0,2\pi)$, $\textbf{Q}_0=(Q_{0,x},Q_{0,y},Q_{0,z})^T$ is an arbitrary real three dimentional vector, $n_A=(N_1+N_2)/V, n_M=N_M/V$ with system volume $V$ are the total atomic density and molecular density respectively.
	
	Furthermore, we derive the free energy density by substituting the above ansatz~(\ref{eq_ouransatz}) to the Hamiltonian density~(\ref{eq_mainH})
	\begin{equation}\label{eq_F}
	\begin{split}
	F/V=&\sum_{\sigma=1,2}\frac{Q^{2}}{2m}n_{\sigma}-\epsilon_{0}n_{M}-\mu(n_{1}+n_{2}+2n_{M}-n)\\
	&+\frac{\overline{g}}{2}n_A\sqrt{n_M}\sin{2\chi_A}[e^{-i\theta_M}(\cos{\chi_M},-i\sin{\chi_M},0)\\
	&\cdot\textbf{Q}_0+h.c]-Z(n_{1}-n_{2}),
	\end{split}
	\end{equation}
	where $n_{1,2}=N_{1,2}/V$, $\mu$ and $Z$ are the Lagrange multipliers set for the conservations of the total particle number and atom-number difference. For simplicity, we only consider a nonpolarized situation in this paper, i.e. $n_1=n_2=n_A/2$, and fix the total particle number. The free energy density does not depend on $\theta, \theta_{x,z}, \overline{\theta}_{x,y,z}$. To minimize the free energy, we obtain the optimal values for the parameters: $\theta_M=0, \chi_A=\pi/4, \chi_M=0, Q_{0,x}=|\textbf{Q}|, Q_{0,y,z}=0$, from which we can see that $\Phi$ is real and parallel to $\textbf{Q}$ by setting $\theta=0$. To be more convenient, we set $\overline{\theta}_{y}=\pi/2, \overline{\theta}_{x,z}=0$ so that $\textbf{Q}$ and $\Phi$ are aligned to $z$ direction. Without loss of generosity, we choose $\overline{g}$ to be negative (if $\overline{g}>0$, $\textbf{Q}$ will be opposite to $\Phi$, however, it gives us the same phases and LHY corrections as we obtain below). Gross-Pitaevskii (GP) equations can be derived from the free energy density formula, and we obtain the optimized solutions to minimize the free energy.
	
	Similar to previous literatures~\cite{radzihovsky2009p, pwave2011Sungsoo}, the ground state phase diagram of our model is also divided into three phases for different detuning $\epsilon_0$, where the ground state phases are listed in the Tab.~\ref{tab_PD}. Here ASF refers to the atomic superfluid phase, where only atomic condensates exist. Note that there is no superfluidity here due to the absence of background atom-atom interaction, where the name of phase is only taken to be consistent with previous convention~\cite{radzihovsky2009p, pwave2011Sungsoo}. AMSF refers to the atomic-molecular superfluid phase, where atom and molecular condensates are present in the same phase. MSF with only molecular condensate is the molecular superfluid phase.
	
	In ASF phase, the condensate in both atomic species stays stationary due to vanishing $\textbf{Q}$ and the two condensates do not interact. The atomic chemical potential remains zero. In AMSF phase, the rotation and time-reversal symmetries are all broken due to the finite-momentum condensates. The $SO(3)$ rotation symmetry is spontaneously broken into $SO(2)$ symmetry. In MSF phase,  although the atomic condensates density is zero, we still have non-zero $\textbf{Q}$. This results in a MSF excitation spectrum translated in momentum space by $\textbf{Q}$ as we will see in section IV.

	From Tab.~\ref{tab_PD}, we can also find the total energy $E_0$ is proportional to particle number $N=nV$ in phases ASF and MSF, which is due to the lack of background atom-atom and molecule-molecule interactions in these phases, respectively. It means the total energy $E_0$ is constant, such that the ground state is stable, for a system with fixed total particle number. However, we can find it is energetically favorable to increase density $n$ to reach lower total energy $E_0$ in AMSF phase. It implies that in this phase the mean-field ground state is unstable and tends to collapse into a state with smaller volume but large particle density when the total particle number is fixed. The instability of the ground state in AMSF phase also manifests itself in the fact that the excitation mode becomes complex in the long-wavelength limit $k \to 0$~\cite{lhyimag}. It will be shown the ground state collapses into a small droplet after considering LHY correction~\cite{LHYoringinalpaper}. In order to calculate this correction, we need to analyze the Bogoliubov excitation spectrum at first.
	
	\begin{figure*}
		\includegraphics[width=16.5cm]{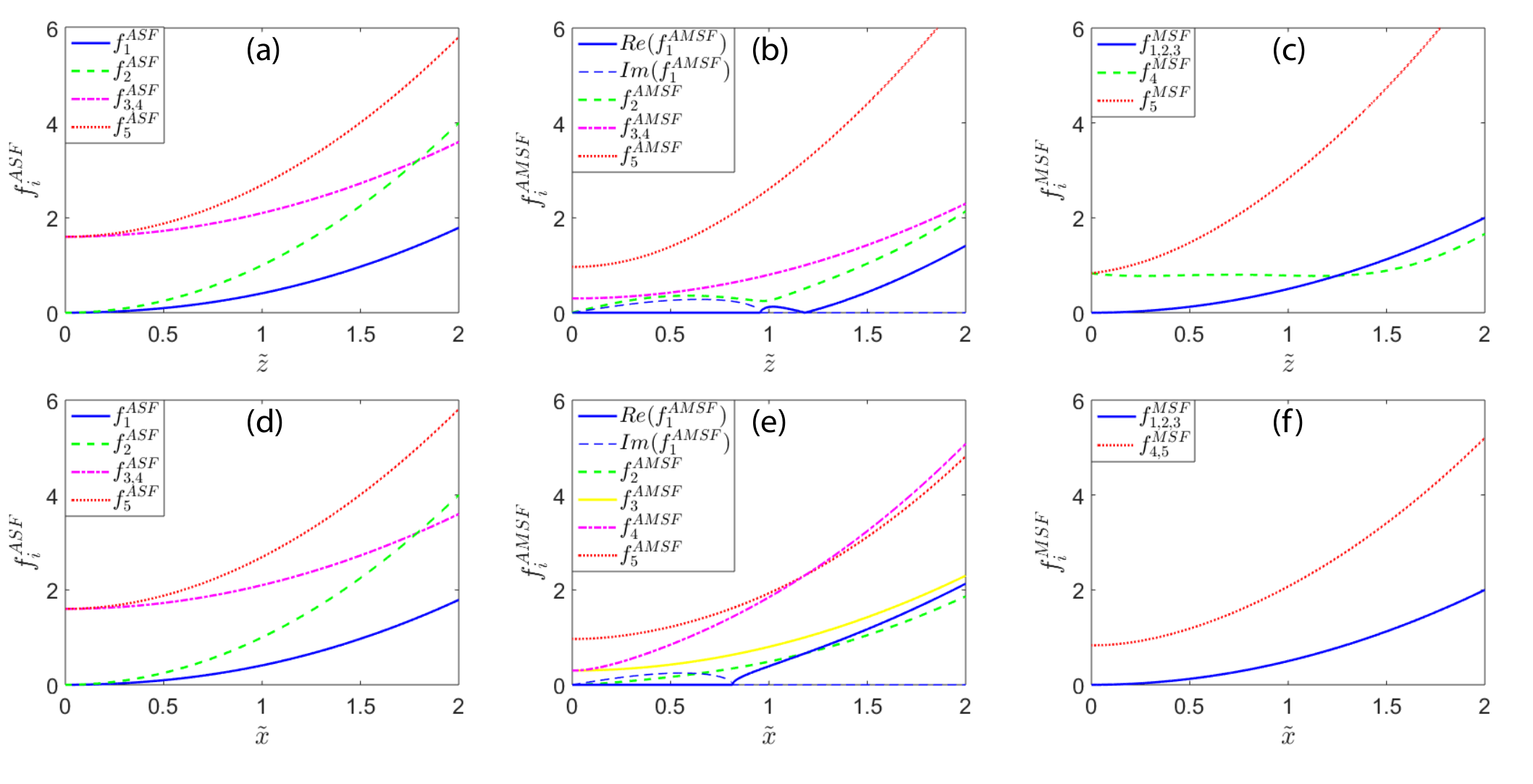}
		\caption{Schematic plot of dimensionless function $f_i$ for $\Delta=-0.4$ (a)(d), $\Delta=0.1$ (b)(e) and $\Delta=0.4$ (c)(f), which is inside ASF, AMSF and MSF phases, respectively. We can find the low-energy modes become imaginary in phase AMSF, which arises from the instability of the mean-field ground state. Here $\tilde{r}=\sqrt{\tilde{x}^2+\tilde{y}^2+\tilde{z}^2}$ represents the distance from the momentum-space origin. For convenience, arbitrary-unit convention is used in this paper. }
		\label{fig:BogoSpectr}
	\end{figure*}

	\section{Bogoliubov excitation spectrum}
	We will study the Bogoliubov excitation spectrum in this section. Following Bogoliubov's theory~\cite{bogolyubov1947theory,pethick2008bose}, we expand the atomic and molecular fields around the ground-state mean fields,
	\begin{equation}\label{eq_bogoe}
	\hat{\psi}_{\sigma}=\Psi_\sigma+\delta \hat{\psi}_{\sigma},\quad \hat{\phi}_i=\Phi_i+\delta \hat{\phi}_i,
	\end{equation}
	with the fluctuation fields $\delta \hat{\psi}_{\sigma}$ and $\delta \hat{\phi}_i$. For convenience, we furthermore expand $\delta \hat{\psi}_{\sigma}$ and $\delta \hat{\phi}_i$ with the Fourier transformation
	\begin{equation}\label{eq_rme}
	\delta \hat{\psi}_{\sigma}=\frac{1}{\sqrt{V}}\sum_{\textbf{k}}\delta\hat{a}_{\sigma, \textbf{k}}e^{-i\textbf{k} \cdot \textbf{r}},\quad \delta \hat{\phi}_{i}=\frac{1}{\sqrt{V}}\sum_{\textbf{k}}\delta\hat{b}_{i, \textbf{k}}e^{-i\textbf{k} \cdot \textbf{r}},
	\end{equation}
	where $\delta\hat{a}_{\sigma, \textbf{k}}$ and $\delta\hat{b}_{i, \textbf{k}}$ are the corresponding quantum fluctuation fields in momentum space. Substituting Eqs.~(\ref{eq_bogoe}) and (\ref{eq_rme}) into Eq.~(\ref{eq_mainH}) and keeping only the second-order terms (the first-order terms are vanished due to the saddle-point solution and higher-order terms will be neglected), we can derive the Bogoliubov Hamiltonian. The Bogoliubov excitation spectrum can be extracted by diagonalizing Bogoliubov Hamiltonian.
	
	\subsection{ASF phase}
	This phase has only atomic condensates, i.e. $n_1=n_2=n/2$, $n_M=0$ and the zero atomic condensates momentum $\textbf{Q}=0$.  The Bogoliubov Hamiltonian can be written as
	\begin{equation}
	\begin{split}
	H_{f}=&\frac{1}{2}\sum_{\textbf{k}}\{\sum_{\sigma=1,2}\varepsilon_{\sigma,\textbf{k}+\textbf{Q}_{\sigma}}\delta\hat{a}_{\sigma,\textbf{k}+\textbf{Q}_{\sigma}}^{\dagger}\delta\hat{a}_{\sigma,\textbf{k}+\textbf{Q}_{\sigma}}\\
	&+\sum_{i}\omega_{i,k}\delta\hat{b}_{i,k}^{\dagger}\delta\hat{b}_{i,k}-2\sum_{\sigma,i}\alpha_{i,\overline{\sigma},\textbf{k}}\delta\hat{b}_{i,k}^{\dagger}\delta\hat{a}_{\sigma,\textbf{k}+\textbf{Q}_{\sigma}}\\
	&+h.c.\},
	\end{split}
	\end{equation}
	\noindent
	where $\overline{\sigma}=3-\sigma, \sigma=1,2$, and the parameters are given as
	\begin{equation}
	\begin{split}
	&\varepsilon_{\sigma,\textbf{k}}=\epsilon_k=\frac{k^2}{2m},\quad\omega_{i,\textbf{k}}=\frac{1}{2}\epsilon_k+\frac{1}{2}\overline{g}^2mn,\\
	&\alpha_{i,\sigma,\textbf{k}}=\mp \frac{1}{2\sqrt{2}}\overline{g}\sqrt{n}k_i.
	\end{split}
	\end{equation}
	The corresponding Bogoliubov excitation spectrum is given by
	\begin{equation}
	E_i^{ASF}=\frac{1}{4}\overline{g}^2mn f^{ASF}_i(\tilde{x},\tilde{y},\tilde{z}), i=1,\cdots,5
	\end{equation}
	where $f_i^{ASF}$ is a dimensionless function, and $\tilde{x}=\frac{k_x}{\overline{g}m\sqrt{n}},\tilde{y}=\frac{k_y}{\overline{g}m\sqrt{n}},\tilde{z}=\frac{k_z}{\overline{g}m\sqrt{n}}$.
	
	We show $f^{ASF}_i$ along the radial direction in Fig.~\ref{fig:BogoSpectr}(a) and (d). The spectrum is symmetric in all directions and has two gapless atomic modes. The quadratic dispersions of gapless mode are due to the absence of atom-atom interaction.
	
	\subsection{AMSF phase}
	In AMSF phase, particles are condensed into both the atomic and molecular channels, and the atomic condensates carry opposite finite momentums. The directions of atomic momentum $\textbf{Q}$ and molecular condensates order parameter $\Phi=(\Phi_x,\Phi_y,\Phi_z)$ are parallel in mean-field ground state, where the direction of $\Phi$ is defined by the it three spatial components. For convenience, we build the coordinate so that this direction is aligned along $z$ axis. The Bogoliubov Hamiltonian is written as
	\begin{equation}
	\begin{split} H_f=&\sum_{\textbf{k}}\{\sum_{\sigma=1,2}\frac{1}{2}\varepsilon_{\sigma,\textbf{k}+\textbf{Q}_\sigma}\delta\hat{a}_{\sigma,\textbf{k}+\textbf{Q}_\sigma}^\dagger \delta\hat{a}_{\sigma,\textbf{k}+\textbf{Q}_\sigma}\\
	&+\sum_{i} \frac{1}{2}\omega_{i,k} \delta\hat{b}_{i,k}^\dagger \delta\hat{b}_{i,k}+t_{\textbf{k}+\textbf{Q}} \delta\hat{a}_{1,\textbf{k}+\textbf{Q}} \delta\hat{a}_{2,-\textbf{k}-\textbf{Q}}\\
	&-\sum_{\sigma,i} \alpha_{i,\overline{\sigma},\textbf{k}} \delta\hat{b}_{i,k}^\dagger \delta\hat{a}_{\sigma, \textbf{k}+\textbf{Q}_\sigma} + h.c.\},
	\end{split}
	\end{equation}
	where the parameters are given by
	\begin{equation}
	\begin{split}
	&\varepsilon_{\sigma,\textbf{k}}=\epsilon_k+\frac{1}{8}\overline{g}^2mn + \frac{1}{4}\epsilon_0,\quad\omega_{i,\textbf{k}}=\frac{1}{2}\epsilon_k + \frac{1}{4}\overline{g}^2mn - \frac{1}{2}\epsilon_0,\\
	&t_{\textbf{k}}=-\overline{g}\sum_{i}\Phi_i^\ast k_i,\quad\alpha_{i,\overline{\sigma},\textbf{k}}=\pm \overline{g}\sqrt{n_\sigma}(Q_{\sigma,i}-k_i/2),
	\end{split}
	\end{equation}
	with $\epsilon_k=\frac{k^2}{2m}$, $\sigma=1,2$ (correspondingly $\bar{\sigma}=2,1$), $\textbf{Q}_{1}=\textbf{Q}$ and $\textbf{Q}_{2}=-\textbf{Q}$. The Bogoliubov excitation spectrum can be written as
	\begin{equation}
	E_i^{AMSF}=\frac{1}{4}\overline{g}^2mn f_i^{AMSF}(\tilde{x},\tilde{y},\tilde{z},\Delta), i=1,\cdots,5
	\end{equation}
	where $f_i^{AMSF}$ is a dimensionless function, $\tilde{x}=\frac{k_x}{\overline{g}m\sqrt{n}},\tilde{y}=\frac{k_y}{\overline{g}m\sqrt{n}},\tilde{z}=\frac{k_z}{\overline{g}m\sqrt{n}}$ and $\Delta=\frac{\epsilon_0}{2\overline{g}^2mn}$.
	
	The schematic plots of $f_i^{AMSF}$ are shown in Fig.~\ref{fig:BogoSpectr}(b) and (e) along $z$ and $x$ directions respectively. As we can see from the two figures, the blue-dashed curve shows imaginary mode consistent with the instability of the mean-field ground state~\cite{lhyimag}, which is absent when the ground state is stable~\cite{radzihovsky2009p,pwave2011Sungsoo}. Actually, the true ground state is lost due to the homogeneous assumption (the system with finite particle number will collapse into a droplet shape that breaks the spatial translation symmetry) and the absence of LHY correction. On the other hand, the inverse of the largest momentum carried by imaginary modes is expected to be comparable with the size of droplet~\cite{lhyimag}. The minima on the blue-solid curve in Fig.~\ref{fig:BogoSpectr}(b) corresponds to the nonvanishing momentum $2\textbf{Q}$ in AMSF phase, where the atomic condensates locate. That the
	spectrum softens to zero at $k_z=2Q$ implies our ansatz correctly captures the feature of the ground state.
	
	\subsection{MSF phase}
	In this phase, we have $n_M=n/2$ and $n_1=n_2=0$. The Bogoliubov Hamiltonian is given as
	\begin{equation}
	\begin{split}
	H_f=&\sum_{\textbf{k}}\{\sum_{\sigma=1,2}\frac{1}{2}\varepsilon_{\sigma,\textbf{k}+\textbf{Q}_\sigma}\delta\hat{a}_{\sigma,\textbf{k}+\textbf{Q}_\sigma}^\dagger \delta\hat{a}_{\sigma,\textbf{k}+\textbf{Q}_\sigma}\\
	&+t_{\textbf{k}+\textbf{Q}} \delta\hat{a}_{1,\textbf{k}+\textbf{Q}} \delta\hat{a}_{2,-\textbf{k}-\textbf{Q}}\\
	&+\sum_{i} \frac{1}{2}\omega_{i,k} \delta\hat{a}_{i,k}^\dagger \delta\hat{a}_{i,k}
	+ h.c.
	\},
	\end{split}
	\end{equation}
	where $\varepsilon_{\sigma,\textbf{k}}=\epsilon_k+\frac{1}{4}\overline{g}^2 m n,\quad \omega_{i,\textbf{k}}=\frac{1}{2}\epsilon_k$ and $\quad t_{\textbf{k}}=-\overline{g}\sum_{i}\Phi_i^\ast k_i$.

	Fortunately, we can derive analytical formulas for the excitation modes in this phase, i.e.
	\begin{equation}
	\begin{split}
	&E_1^{MSF}=\frac{1}{4}\overline{g}^2mn\sqrt{\tilde{r}^2(2+\tilde{r}^2-2\sqrt{2}\tilde{r}\cos{\gamma})},\\
	&E_2^{MSF}=\frac{1}{4}\overline{g}^2mn\sqrt{\tilde{r}^2(2+\tilde{r}^2+2\sqrt{2}\tilde{r}\cos{\gamma})},\\
	&E_{3,4,5}^{MSF}=\frac{1}{8}\overline{g}^2mn\tilde{r}^2.
	\end{split}
	\end{equation}
	where $\tilde{r}^2=(k_x^2+k_y^2+k_z^2)/\overline{g}^2m^2n$ and $\gamma$ is the angle between $z$ axis and unit vector $\hat{\textbf{k}}$ as we have aligned $\hat{\textbf{Q}}$ along $z$. Similar to what we defined in ASF and AMSF phases, we rewrite the excitation modes in this formula
	\begin{equation}
	E_i^{MSF}=\frac{1}{4}\overline{g}^2mn f_i^{MSF}(\tilde{x},\tilde{y},\tilde{z},\Delta), i=1,\cdots,5
	\end{equation}
	Fig.~\ref{fig:BogoSpectr}(c) and (f) are the corresponding $f_i^{MSF}$ along $z$ and $x$ directions. In Fig.~\ref{fig:BogoSpectr}(b), the red dotted and green dashed curves are the two atom modes respectively, and the minima on green dashed curve corresponds to the nonvanishing momentum $2\textbf{Q}$. The blue curve denotes the triply-degenerated molecule modes. In Fig.~\ref{fig:BogoSpectr}(f), the red curve denotes the doubly-degenerated atom modes and blue curve denotes the triply-degenerated molecule modes.

	\begin{figure}
		\includegraphics[width=8cm]{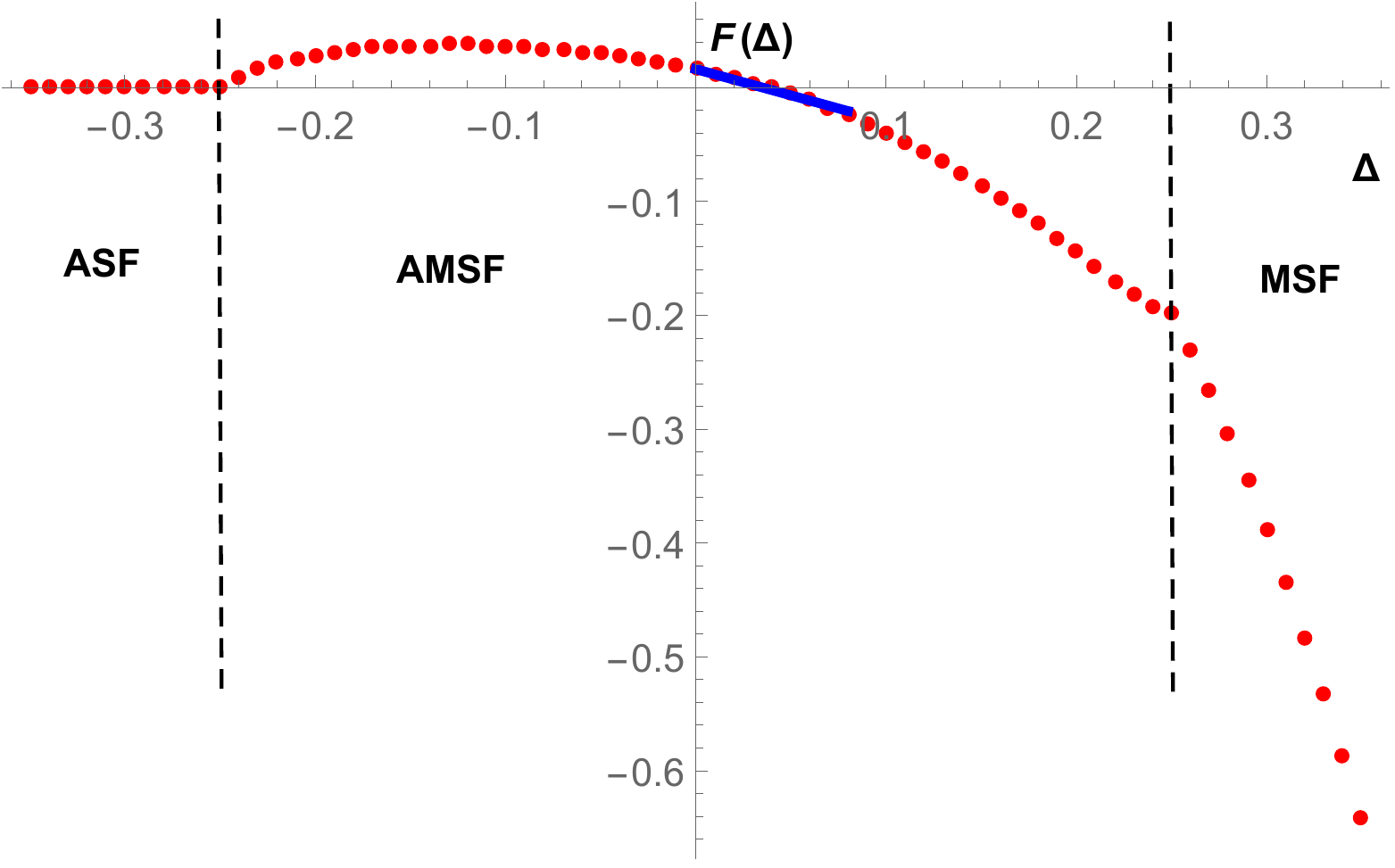}
		\caption{Schematic plot of $F(\Delta)$. The blue solid line is a linearized approximation for the regime with a stabilized particle number density.}
		\label{fig:F(Delta)}
	\end{figure}

	\begin{figure*}
		\includegraphics[width=17.5cm]{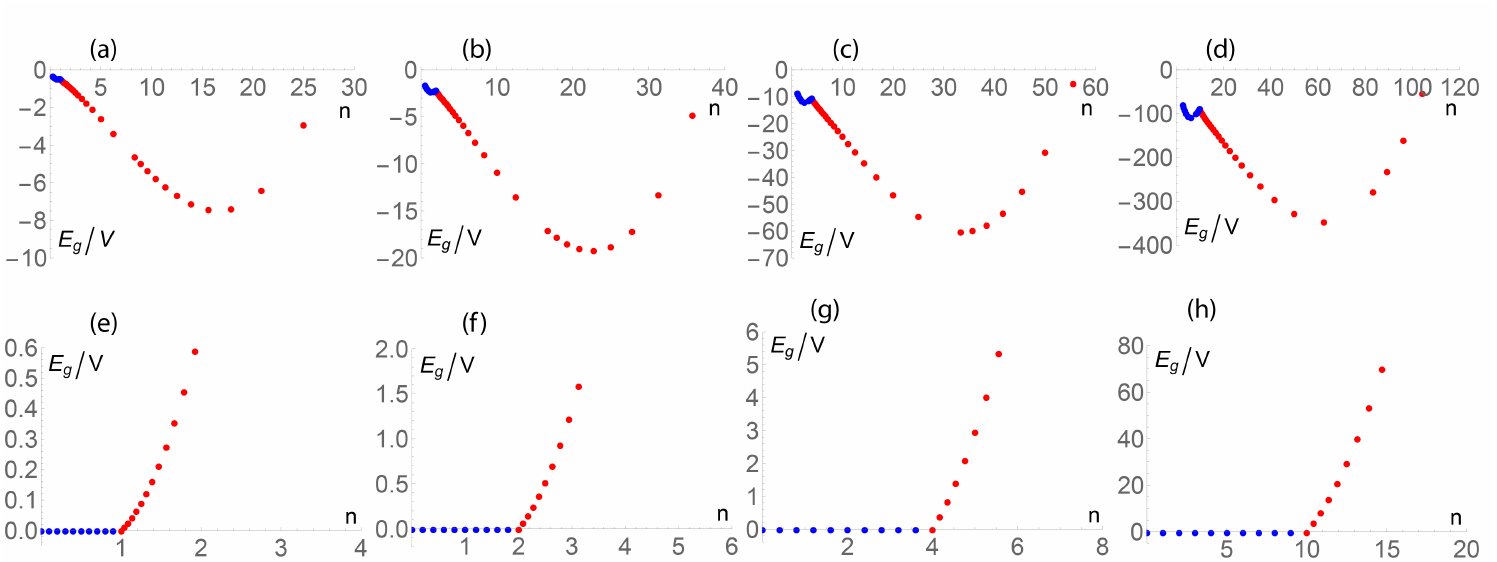}
		\caption{Total ground-state energy density versus total number density for different detuning: $\epsilon_0=0.5$ (a), $1$ (b), $2$ (c), $5$ (d), $-0.5$ (e), $-1$ (f), $-2$ (g), $-5$ (h), and here we set $\overline{g}=-1,m=1$. In subfigures with $\epsilon_0>0$, the blue (red) dots repersent the energies in MSF (AMSF) phase.  The minimum energy density is presented with a finite number density after we introduce the LHY correction and lies in the AMSF phase. In subfigures with $\epsilon_0<0$, the blue (red) dots repersent the energies in ASF (AMSF) phase.  }
		\label{fig:Evsn}
	\end{figure*}

	\section{LHY correction}
	The LHY correction is the leading-order correction of quantum fluctuation. It is composed of Bogoliubov excitation energies, commutation energies which appear due to the commutation relations of Nambu basis, and energy correction due to the interaction renormalization. Here the interaction renormalization is employed to remove the energy divergence arising in collecting the energy of quantum fluctuation~\cite{LHYoringinalpaper}. Let us review the renormalization procedure before going ahead.
	
	To remove the divergence appears in the calculation of LHY correction, we need to renormalize the interaction parameter $\overline{g}$ and detuning $\epsilon_0$~\cite{LHYoringinalpaper,lhyimag}. The two body T matrix for p wave interaction is given by~\cite{radzihovskypwave}
	\begin{equation}\label{eq_TMatr}
	\begin{split}
	&-iT_{\textbf{k},\textbf{k}'}^{(l_z)}(k)=D^{(0)}(k)(-i\overline{g})^2 k^2 Y_{1,l_z}(\hat{\textbf{k}}) Y_{1,l_z}^\ast(\hat{\textbf{k}}')\\
	&+
	D^{(0)2}(k)(-i\overline{g})^4 2\Pi_{l_z}(k) k^2 Y_{1,l_z}(\hat{\textbf{k}}) Y_{1,l_z}^\ast(\hat{\textbf{k}}')+...\\
	&=D(k)(-i\overline{g})^2 k^2 Y_{1,l_z}(\hat{\textbf{k}}) Y_{1,l_z}^\ast(\hat{\textbf{k}}').
	\end{split}
	\end{equation}
	where the index $l_z$ denotes different interacting channels $l_z=-1,0,1$. $Y_{1,l_z}(\hat{\textbf{k}})$ is the $l_z$-th channel of the first order spherical harmonics. $D^{(0)}(k)$ is the p-wave scattering propagator, and $\Pi_{l_z}(k)$ is the polarization bubble for channel $l_z$, which are given by
	\begin{equation}
	D^{(0)}(k)=\frac{i}{k^2/m+\epsilon_0+i0+},
	\end{equation}
	and
	\begin{equation}
	\Pi_{l_z}(k)=\int \frac{d^3p}{(2\pi)^3}\frac{ip^2 |Y_{1,l_z}(\hat{\textbf{p}})|^2}{k^2/m-p^2/m+i0+}.
	\end{equation}
	\noindent
	Using Eq. (\ref{eq_TMatr}), we yield
	\begin{equation}\label{eq_DP}
	D^{-1}(k)=[D^{(0)}(k)]^{-1}-(-i\overline{g})^2 \Pi_{l_z}(k).
	\end{equation}
	Comparing $k^0$ term and $k^2$ term on both sides of Eq. (\ref{eq_DP}), we obtain the renormalization relations~\cite{pwaverenpaper},
	\begin{equation}
	\frac{\tilde{\epsilon}_0}{\tilde{\overline{g}}^2}=\frac{\epsilon_0}{\overline{g}^2}+\int \frac{d^3p}{(2\pi)^3}m|Y_{1,l_z}(\hat{\textbf{p}})|^2,
	\end{equation}
	and
	\begin{equation}
	\frac{1}{\tilde{\overline{g}}^2}=\frac{1}{\overline{g}^2}-\int\frac{d^3p}{(2\pi)^3}m^2\frac{|Y_{1,l_z}(\hat{\textbf{p}})|^2}{p^2},
	\end{equation}
	\noindent
	where $\tilde{\epsilon}_0$ and  $\tilde{\overline{g}}$ are the renormalized detuning and p-wave interacting strength respectively.
	
	Applying these renormalization relations into the ground state energy in different phases, one obtains the renormalized mean-field ground state energies
	\begin{equation}
	E_{0,r}^{ASF}/V=0,
	\end{equation}
	\begin{equation}
	E_{0,r}^{AMSF}/V=\int \frac{d^3k}{(2\pi)^3}\frac{1}{12}\overline{g}^2mn+\frac{1}{6}\epsilon_0+\frac{\overline{g}^4m^2n^2-4\epsilon_0^2}{72k^2},
	\end{equation}
	and
	\begin{equation}
	E_{0,r}^{MSF}/V=\int \frac{d^3k}{(2\pi)^3}\frac{1}{6}\overline{g}^2mn+\frac{\overline{g}^2mn-2\epsilon_0}{12k^2}.
	\end{equation}
	
	From the analysis of Bogoliubov spectrum and interaction renormalization, we obtain the LHY correction densities in different phases,
	\begin{equation}
	\begin{split}
	E_{LHY}^{ASF}/V=&-\overline{g}^5m^4n^{2.5}\int\frac{d^3\tilde{r}}{(2\pi)^3}\sum_{i=1}^5 \frac{1}{4}f_i^{ASF}\\
	&-\frac{7}{8}\tilde{r}^2-\frac{3}{8},
	\end{split}
	\end{equation}
	\begin{equation}
	\begin{split}
	E_{LHY}^{AMSF}/V=&-\overline{g}^5m^4n^{2.5}\int\frac{d^3\tilde{r}}{(2\pi)^3}\sum_{i=1}^5 \frac{1}{4}f_i^{AMSF}\\
	&-\frac{7}{8}\tilde{r}^2-\frac{13}{24}+\frac{5}{6}\Delta+\frac{1-16\Delta^2}{72\tilde{r}^2},
	\end{split}
	\end{equation}
	and
	\begin{equation}
	\begin{split}
	E_{LHY}^{MSF}/V=&-\overline{g}^5m^4n^{2.5}\int\frac{d^3\tilde{r}}{(2\pi)^3}\sum_{i=1}^5 \frac{1}{4}f_i^{MSF}\\
	&-
	\frac{7}{8}\tilde{r}^2-\frac{1}{3}+\frac{1-4\Delta}{12\tilde{r}^2}.
	\end{split}
	\end{equation}
	
	Combining the mean-field ground state energy densities and LHY corrected energy densities yields the total ground state energy density $E_g/V=E_{0}/V+E_{LHY}/V$ for different phases as follows,
	\begin{equation}
	E_{g}^{ASF}/V=0,
	\end{equation}
	\begin{equation}\label{eq_EgAMSF}
	\begin{split}
	E_{g}^{AMSF}/V=&-\overline{g}^5m^4n^{2.5}F(\Delta)-\frac{1}{16}\overline{g}^2mn^2\\
	&-\frac{1}{4}\epsilon_0n-\frac{1}{4}\frac{\epsilon_0^2}{\overline{g}^2m},
	\end{split}
	\end{equation}
	and
	\begin{equation}
	E_{g}^{MSF}/V=-\overline{g}^5m^4n^{2.5}F(\Delta)-\frac{1}{2}n\epsilon_0,
	\end{equation}
	\noindent
	where $F(\Delta)$ is depicted in Fig.~\ref{fig:F(Delta)} numerically.
	
\begin{figure}
		\includegraphics[width=8cm]{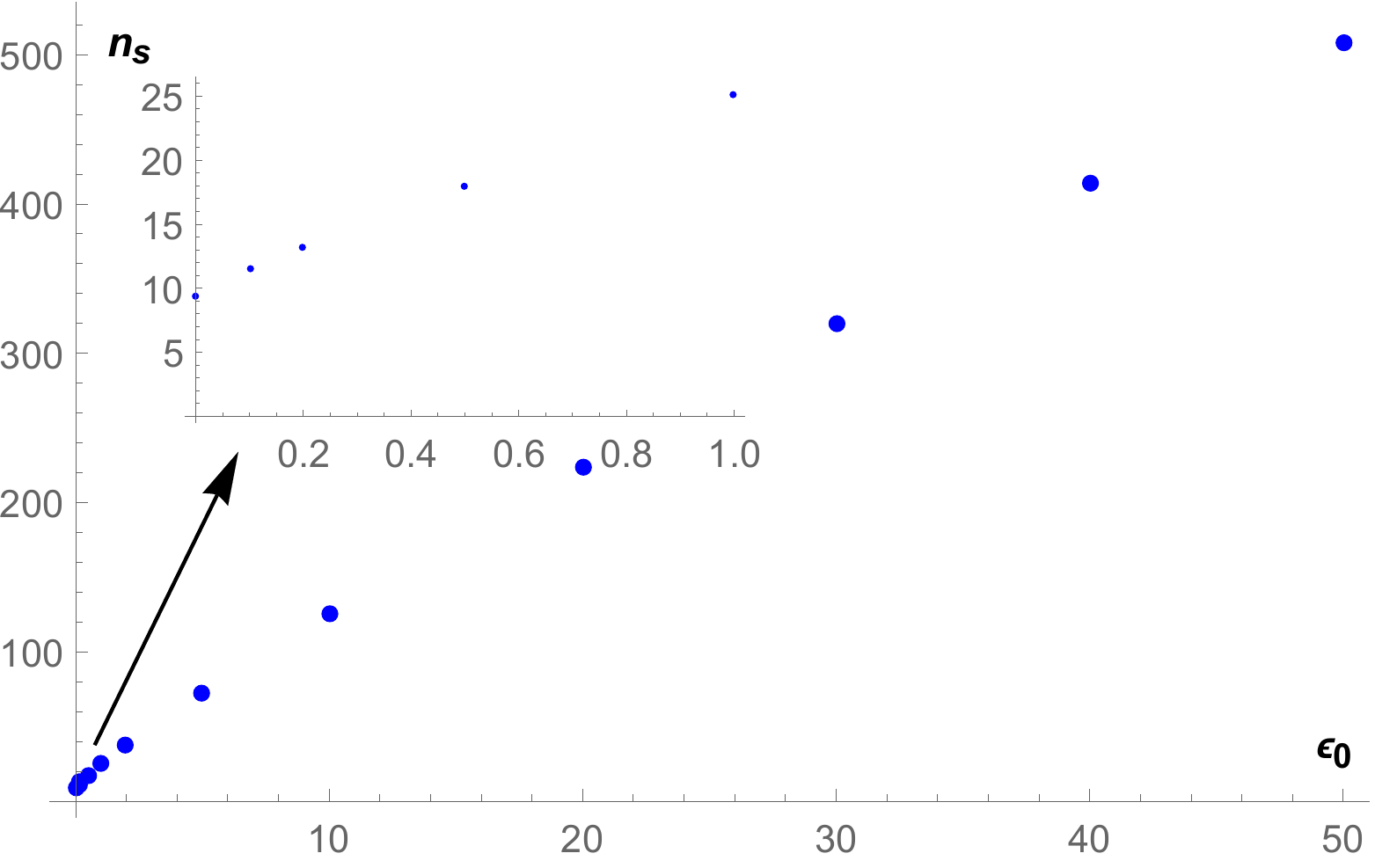}
		\caption{The stabilized density $n_s$ versus detuning $\epsilon_0$ for $\overline{g}=-1,m=1$. The stabilized density $n_s$ is almost proportional to detuning $\epsilon_0$ linearly. As $\epsilon_0$ becomes larger, $\Delta$ converges to $\approx 0.08$.}\label{fig:nvse0new}
	\end{figure}

	We plot the total energy density versus particle density for different detuning in Fig.~\ref{fig:Evsn}. As we can see, for $\epsilon_0>0$, the minimum energy density is well defined, and lies in the AMSF phase. We expect there exists a self-stabilized state at around the minimum. If the particle number is finite, it forms a quantum droplet~\cite{petrov2015quantum}. We also depict the dependence between the particle density of the self-stabilized state $n_s$ and the detuning $\epsilon_0$ in Fig.~\ref{fig:nvse0new}.  It is shown the stabilized density is almost linearly proportional to $\epsilon_0$. However, if $\epsilon_0<0$, the energy density is degenerated inside ASF phase, but it can be broken by introducing an atom-atom s-wave interaction. Typically, the atom-atom s-wave interaction is repulsive and the corresponding LHY correction is also positive~\cite{LHYoringinalpaper}. Therefore, the lowest energy density lies at $n=0$ inside ASF phase. For this reason, we do not expect a self-stabilized state when the detuning $\epsilon_0<0$.
	
\begin{figure}
		\includegraphics[width=8cm]{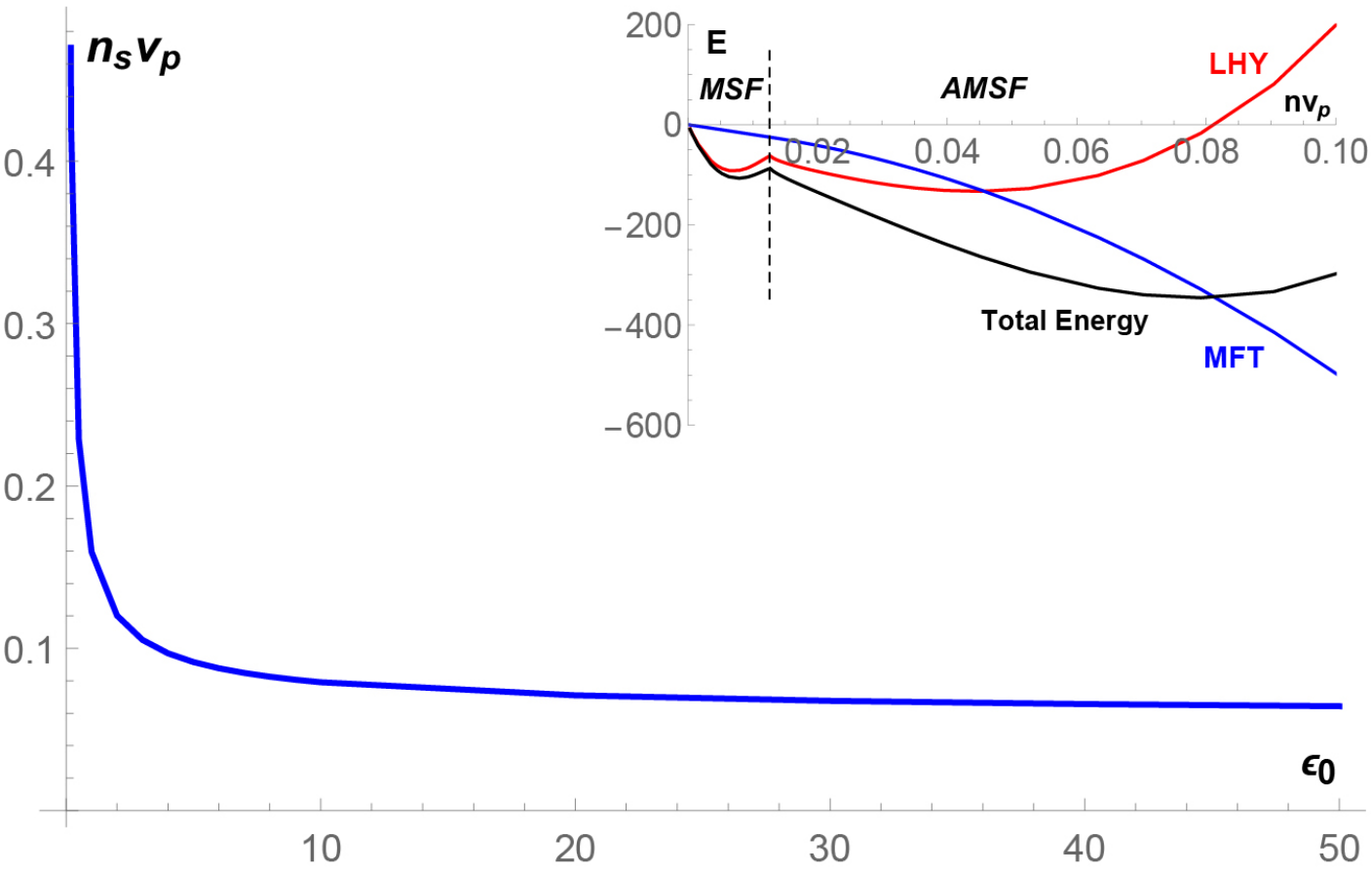}
		\caption{Relation between dilutness and detuning for $\overline{g}=-1, m=1$. As detuning approaches zero, the diluteness tends to diverge, which may indicate that higher order corrections beside MFT and LHY are needed. But for a large detuning regime, the mixture is dilute, so that it is reasonable to charaterize our model with only first order beyond-mean-field calculation. The inset shows the energy comparison for different diluteness, as we set $\overline{g}=-1, m=1, \epsilon_0=5$. The lowest total energy is ensured to appear in the dilute regime.}
		\label{fig:Diluteness}
	\end{figure}

	The diluteness of p-wave interacting gas can be characterized by the product between the particle density and the scattering volume $v_p$~\cite{pwaverenpaper, pwavescatvol}, i.e.
	$nv_p=\overline{g}^2mn/(16\pi^2\epsilon_0)$. Therefore, we can rewrite the ground-state energy given by mean-field theory (MFT) and the LHY correction in terms of the diluteness as
	\begin{equation}
	E^{MFT}_{AMSF}=-\frac{\epsilon_0^2}{64\overline{g}^2m}(32\pi^2nv_p+4)^2,
	\end{equation}
	\begin{equation}
	E^{MFT}_{MSF}=-\frac{\epsilon_0^2}{4\overline{g}^2m}32\pi^2nv_p,
	\end{equation}
	and
	\begin{equation}
	E^{LHY}=m^{1.5}(\epsilon_0/2)^{2.5}(32\pi^2n v_p)^{2.5}F(\frac{1}{32\pi^{2}nv_p}).
	\end{equation}
	The diluteness of the self-stabilized state with respect to detuning is shown in Fig.~\ref{fig:Diluteness}. As detuning approaches zero, the diluteness tends to diverge, which may indicate that higher order corrections beside MFT and LHY are needed. But for a large detuning regime, the mixture is dilute, so that it is reasonable to characterize our model with only first order beyond-mean-field calculation.
	
	\section{Quantum Droplets}
	According to the above analysis, we find the mean-field collapsing state becomes self-stabilized after considering beyond-mean-field correction. This self-stabilized state forms a quantum droplet when particle number is finite~\cite{petrov2015quantum}. To figure out the density distribution of the quantum droplet, we will derive an effective theory to characterize the density profile. Here we employ function $\xi(\textbf{r})$ to characterize the droplet density profile. If the system size is infinite, we have solution $\xi(\textbf{r})=1$ as it should be a uniform gas. However, if the system size is finite, the density profile will be inhomogeneous.
	
	As a qualitative analysis, we will take the local density approximation (LDA). With this approximation, the order parameters can be rewritten as~\cite{lhyimag},
	\begin{equation}\label{eq_dropletansatz1}
	\Psi_{1}=\Psi_{1,\textbf{Q}}e^{-i\textbf{Q} \cdot \textbf{r}},\quad \Psi_{2}=\Psi_{2,-\textbf{Q}}e^{i\textbf{Q} \cdot \textbf{r}},
	\end{equation}
	and $\Phi=\sqrt{n_{s,M}}\xi(\textbf{r})\hat{\textbf{z}}$, where
	\begin{equation}\label{eq_dropletansatz2}
	\Psi_{1,\textbf{Q}}=\sqrt{n_{s,1}}\xi(\textbf{r}),\quad\Psi_{2,-\textbf{Q}}=\sqrt{n_{s,2}}\xi(\textbf{r}).
	\end{equation}
	Here we have chosen $\hat{\textbf{z}}$ direction due to the spontaneous breaking of $SO(3)$ rotation symmetry by $\textbf{Q}=-\overline{g}m\sqrt{n_{s,M}}\xi(\textbf{r})\hat{\textbf{z}}$, where $n_{s,1,2,M}$ are the stabilized densities for the two atomic species and molecule respectively.
	
	\begin{figure}
		\includegraphics[width=9.2cm]{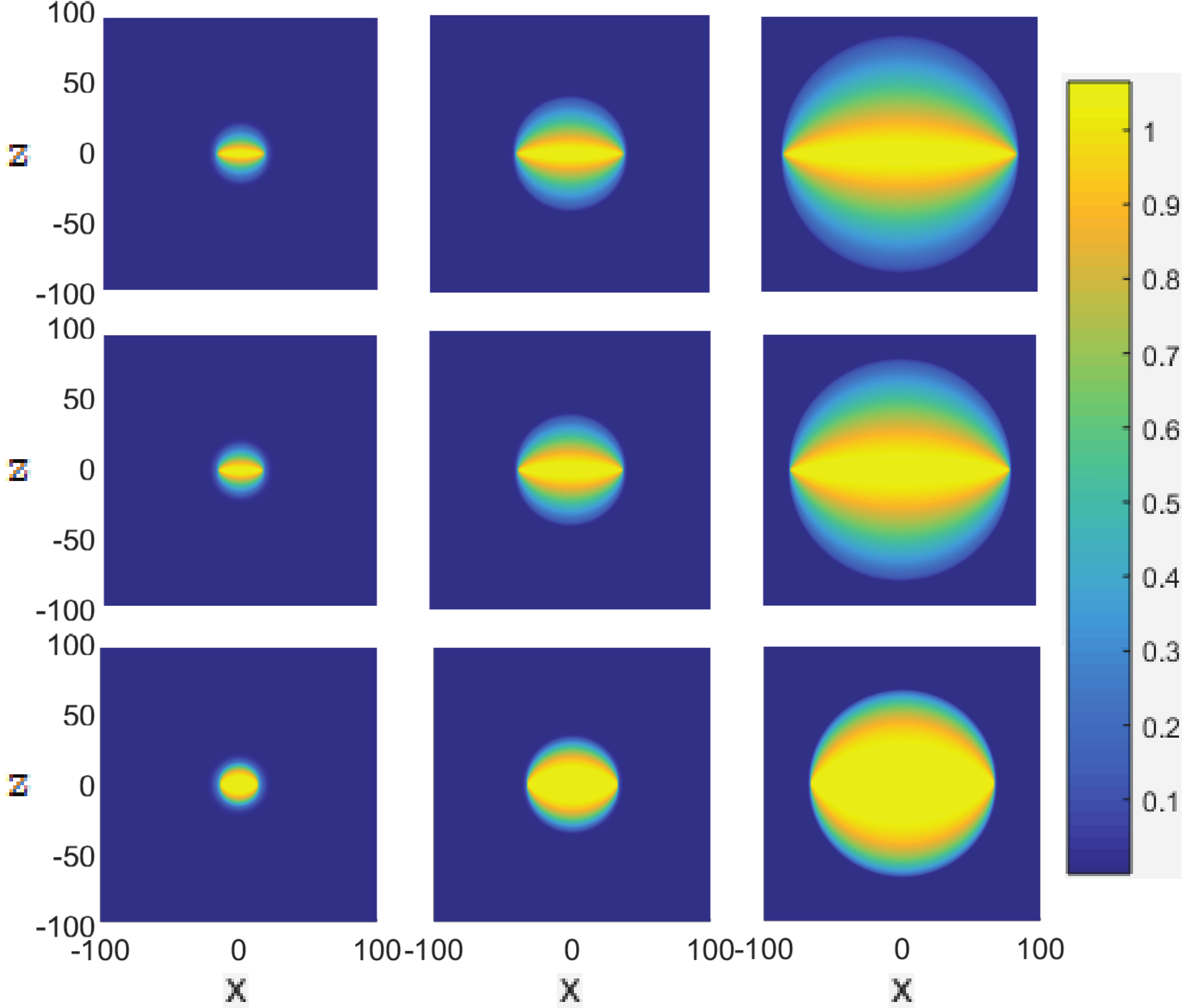}
		\caption{Density profile of droplet under condition $\overline{g}=-1, m=1$. The background color represents $\xi=\sqrt{n(\textbf{r})/n_s}$, where $n(\textbf{r})$ is density at different locations and $n_s$ is the stabilized density. The $x$ axis and $y$ axis for each subfigure label the $x$ direction and $z$ direction in real space. The detunings from the top row to the bottom row are $\epsilon_0=0.5,5,50$ respectively. The normalization factor for $\xi$ from the left column to the right column are $N/n_s=10^4,10^5,10^6$ respectively. When the particle number grows large enough with $n_s$ fixed, it breaks $SO(3)$ symmetry clearly. As the dutuning grows smaller and deep inside the AMSF phase, the droplet is more and more reduced along $z$ axis.}
		\label{fig:dropletshape}
	\end{figure}
	
	To access the analytical form of effective Hamiltonian, an approximative form of $F(\Delta)$ at around the stable point is considered. For $\epsilon_0>0$, we find a linearized formula for $F(\Delta)$, which captures its behavior at around the minimum energy density inside the AMSF phase ($0<\Delta<\approx0.08$) [see Fig.~\ref{fig:F(Delta)}]. It is written as
	\begin{equation}
	F(\Delta) \approx -0.460333\Delta+0.01624807.
	\end{equation}
	According to Eq. (\ref{eq_EgAMSF}), the approximated total ground-state energy in AMSF phase is given by
	\begin{equation}
	\begin{split}
	E_{g}^{AMSF}/V=&-\frac{1}{16}\overline{g}^2mn_s^2
	-\frac{1}{4}\epsilon_0n_s-\frac{1}{4}\frac{\epsilon_0^2}{\overline{g}^2m}\\
	&-0.01625\overline{g}^5m^4n_s^{2.5}+0.2302\epsilon_0\overline{g}^3m^3n_s^{1.5}.
	\end{split}
	\end{equation}
	Furthermore, by substituting Eqs. (\ref{eq_dropletansatz1}) and (\ref{eq_dropletansatz2}) to Eq.~(\ref{eq_mainH}), we derive the effective Hamiltonian
	\begin{equation}
	\begin{split}
	\mathcal{H}_{eff}=&\left.
	\overline{g}^2mn_s^2\{[\frac{1}{\overline{g}^2m^2n_s} (\frac{3}{4}\Delta-\frac{5}{16})+(\Delta^2-\frac{1}{16})z^2\xi^2]\xi \nabla^2 \xi\right.\\
	& \left.
	-
	(2\Delta^2+\frac{1}{2}\Delta)\xi^2+0.460333 \Delta \overline{g}^3m^3\sqrt{n_s}\xi^3\right.\\
	& \left.
	+
	(\Delta^2-\frac{1}{16})\xi^4-0.01625\overline{g}^3m^3\sqrt{n_s}\xi^5\}
	\right..
	\end{split}
	\end{equation}
	The chemical potential $\tilde{\mu}$ is fixed by the normalization condition $\int d^3r |\xi|^2=N/n_s$, where $N$ is the total number of particle and $n_s$ is the stabilized total density. The profile function $\xi(\textbf{r})$ is determined by the GP equation
	\begin{equation}
	\begin{split}
	\tilde{\mu}\xi^2=&\left.
	\overline{g}^2mn_s^2\{[\frac{1}{\overline{g}^2m^2n_s} (\frac{3}{4}\Delta-\frac{5}{16})+2(\Delta^2-\frac{1}{16})z^2\xi^2]\xi \nabla^2 \xi\right.\\
	& \left.-(2\Delta^2+\frac{1}{2}\Delta)\xi^2+0.690501 \Delta \overline{g}^3m^3\sqrt{n_s}\xi^3\right.\\
	& \left.+2(\Delta^2-\frac{1}{16})\xi^4-0.0460202\overline{g}^3m^3\sqrt{n_s}\xi^5\}\right.,
	\end{split}
	\end{equation}
	which is derived by minimizing the effective Hamiltonian.

	\begin{figure}
		\includegraphics[width=7cm]{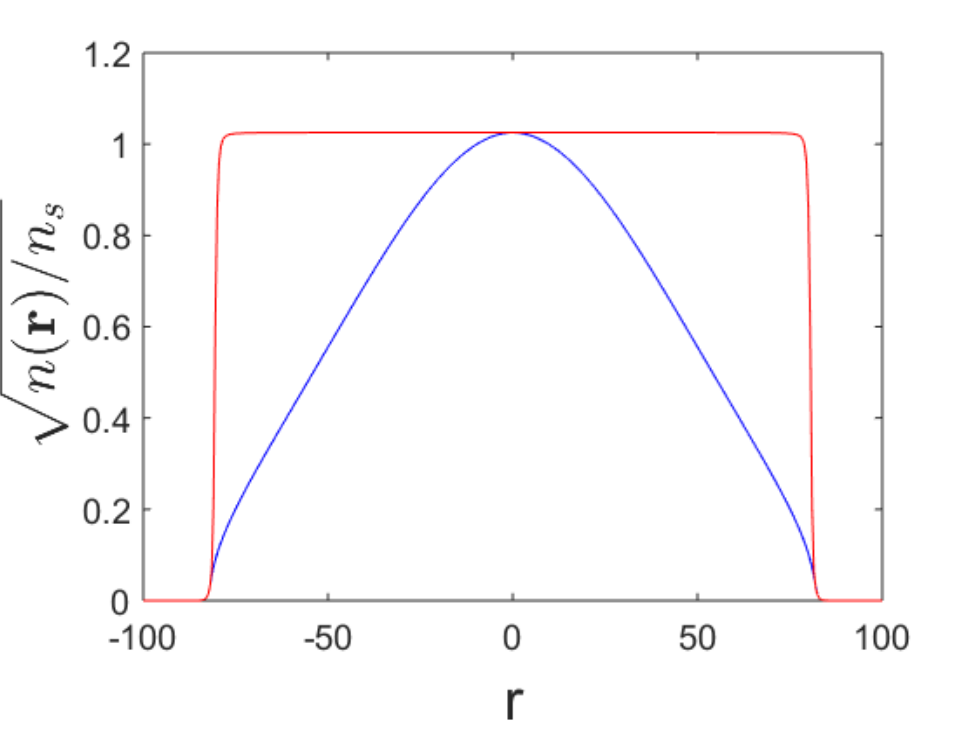}
		\caption{Density profile on the centred lines along $x$ and $z$ directions inside the droplet under condition $\overline{g}=-1, m=1, \epsilon_0=0.5, N/n_s=10^6$. Red curve is the centred line along $x$ direction and blue curve is along $z$ direction. The value on the plateau is almost constant and close to 1.025. If the system size is increased, the height of the plateau will be closer to 1.}\label{fig:crossscan}
	\end{figure}
	
	The above GP equation is solved numerically by using imaginary time evolution method. The solutions for different detuning and particle numbers are shown in Fig.~\ref{fig:dropletshape}. We can find the quantum droplet is typically suppressed in the $z$ direction. The degrees of suppression decreases for a larger $\epsilon_0$.  Hence the droplet looks like a pancake when $N/n_s$ is large enough but $\epsilon_0$ is small (see the upper right subfigure of Fig.~\ref{fig:dropletshape}). We also show the section of the solution where $\overline{g}=-1, m=1, \epsilon_0=0.5, N/n_s=10^6$ in Fig.~\ref{fig:crossscan}. The density is found to suddenly fall to zero in the horizontal directions ($x$ or $y$ directions), while gently decreasing to zero in the $z$ direction. Except for the boundary regime, the profile varies smoothly everywhere, which implies that LDA could qualitatively catch the features of quantum droplet here. In fact, the anisotropy of quantum droplet arises from the spontaneous breaking of SO(3) rotation symmetry by finite-momentum atomic condensates. It is intrinsically different from the anisotropic quantum droplets in the presence of dipolar interaction~\cite{ferrier2016observation,kadau2016observing,chomaz2016quantum,schmitt2016self} or spin-orbit coupling~\cite{cui2018spin}, where the anisotropy arises from external fields. As we can see, the value on the plateau remains almost constant and close to 1, which will be exactly 1 when the system size goes to infinity. Another special feature of quantum droplet here is that the atomic components carry finite momentums due to the breaking of time-reversal symmetry.
	
	\section{Conclusion}
	In this paper, we study the quantum fluctuation correction to the ground states of a p-wave interacting Bose gas. Beginning with the mean-field analysis of the ground states, it is found that the ground states can be divided into three typical phases for different detunings of molecule channel, i.e. the ASF, AMSF and MSF phases, where particles are condensed into only the atomic, both the molecular and atomic, and only the molecular channels, respectively. Particularly, we find the ground state is unstable in phase AMSF. The unstability of the ground state in the phase AMSF also manifests itself in the emergence of imaginary long-wavelength Bogoliubov excitation modes. Furthermore, we calculate the LHY correction with the Bogoliubov excitations. We find the LHY correction can stabilize the ground state in the mean-field-unstable regime. That means that the p-wave interacting Bose gas is self-stabilized at certain density. Finally, we construct an effective Hamiltonian to characterize the ground state of a finite system. By solving the corresponding GP equation, we find self-stabilized quantum-droplet solutions. Unlike the s-wave case, the quantum droplet is anisotropic and carries finite momentums because the spatial rotation and the time-reversal symmetries are spontaneously broken. Although only the interspecies p-wave interaction is considered here, our results could be extended into the case with weak background s-wave interactions and may be observed in systems like $^{85}Rb-^{87}Rb$ Bose mixture~\cite{papp2008tunable,dong2016observation}.
	
	\section*{Acknowledgements}
	The authors are indebted to  Wei Yi, Chao Gao, Jing-Bo Wang and Fang Qin for helpful discussion. This work is supported by the AFOSR Grant No. FA9550-16-1-0006, the MURI-ARO Grant No. W911NF17-1-0323, the ARO Grant No. W911NF-11-1-0230 (Z. L. and W.V. L.), the National Postdoctoral Program for Innovative Talents of China (Grant No. BX201700156), the National Natural Science Foundation of China  (Grant No. 11804221) (J.-S. P.), the Science and Technology Commission of Shanghai Municipality (Grants No.16DZ2260200) and National Natural Science Foundation of China (Grants No.11655002) (J.-S. P. and W.V. L.), and the Overseas Scholar Collaborative Program of NSF of China No. 11429402 sponsored by Peking University (W.V. L.). J.-S. P. also acknowledges the support by Xiong-Jun Liu's group during his visit to Peking University.
	
	\bibliographystyle{apsrev4-1}
	\bibliography{reference}
	
\end{document}